\documentclass[aps,twocolumn,showpacs,superscriptaddress]{revtex4}
\usepackage{graphicx}
\usepackage{amsmath}
\usepackage{amssymb}

\newcommand{\q}[1]{``#1''}
\newcommand{\ub}{U_{\text{B}}}
\newcommand{\ud}{U_{\Delta}}
\newcommand{\vd}{v_{\text{d}}}

\newcommand{\QCSE}{QCSE}
\newcommand{\QW}{QW}
\newcommand{\CQW}{coupled QW}
\newcommand{\PL}{PL}

\newcommand{\ICCD}{intensified CCD}
\newcommand{\BEC}{BEC}

\begin{document}

\title{Dynamics of Long-Living Excitons in Tunable Potential Landscapes}

\author{Andreas~G\"artner}
\thanks{Corresponding author.\\
E-mail: andreas.gaertner@physik.uni-muenchen.de}
\affiliation{Center for Nanoscience and Department f\"ur Physik,
Ludwig-Maximilians-Universit\"at, Geschwister-Scholl-Platz 1, 80539 M\"unchen,
Germany}

\author{Dieter~Schuh}
\affiliation{Institut f\"ur Angewandte und Experimentelle Physik,
Universit\"at Regensburg, 93040 Regensburg, Germany}

\author{J\"org~P.~Kotthaus}
\affiliation{Center for Nanoscience and Department f\"ur Physik,
Ludwig-Maximilians-Universit\"at, Geschwister-Scholl-Platz 1, 80539 M\"unchen,
Germany}

\begin{abstract}
A novel method to experimentally study the dynamics of long-living excitons in
coupled quantum well semiconductor heterostructures is presented.
Lithographically defined top gate electrodes imprint in-plane artificial
potential landscapes for excitons via the quantum confined Stark effect.
Excitons are shuttled laterally in a time-dependent potential landscape defined
by an interdigitated gate structure. Long-range drift exceeding a distance of
$150$\,$\mu$m at an exciton drift velocity $\vd \gtrsim 10^3$\,$\mathrm{m} /
\mathrm{s}$ is observed in a gradient potential formed by a resistive gate
stripe.
\end{abstract}


\maketitle
\section*{Introduction}
In the past few years new material systems such as coupled quantum well (QW)
heterostructures emerged. They allow to host long-living excitons with life
times up to about $\approx$\,30\,$\mu$s~\cite{SnoPRL05}. Being composite bosonic
particles made of an electron and a hole, excitons are expected to show
Bose-Einstein-Condensation ({\BEC}) at low temperatures and at sufficient high
densities~\cite{KelJETP68}. Up to present there is no unambigious experimental
evidence for excitonic {\BEC} \cite{ButNat02b,SnoNat02,RapPRL04,SnoPss03}. One
reason is the lack of spatial control on the exciton gas leading to quick
expansion and dilution of an initially dense exciton cloud. In \CQW\ samples,
exciton confinement has been observed only in intrinsic \q{natural
traps}~\cite{ButNat02a} and in mechanically stressed
configurations~\cite{SnoAPL99}. However, both do not allow an in-situ control of
the trapping potential. A profound understanding of controlling exciton dynamics
is essential to define confinement potentials for {\BEC} experiments. In this
contribution, voltage-tunable potential landscapes for excitons are
experimentally demonstrated enabling spatial and temporal control of exciton
dynamics. Using the quantum confined Stark effect ({\QCSE}), laterally modulated
excitonic potential landscapes are induced in {\CQW}s. Exciton shuttling between
two electrodes in a time-varying lateral potential landscape is demonstrated. In
the last section, long-range exciton drift exceeding 150\,$\mu$m is observed in
a gradient potential defined by a resistive gate, and an estimate for the
exciton drift velocity is given.
\section*{Sample and Experimental Details}
\begin{figure}[t] 
\begin{center}\leavevmode
\includegraphics[width=0.9\linewidth]{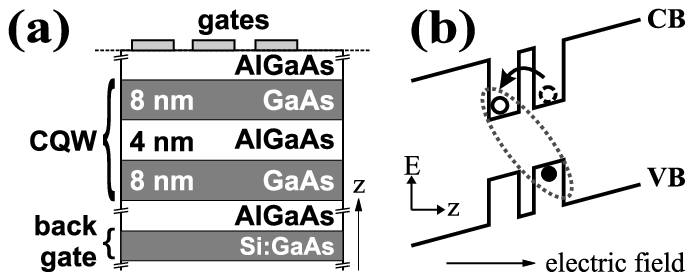}
\end{center}
\caption{
(a) Layout of the heterostructure containing {\CQW}s.
(b) Formation of a spatially indirect exciton (dashed ellipse) in {\CQW}s. The
exciton's life-time and energy are both controllable via an external electric
field applied along the z-axis.
}
\label{fig:indexc}
\end{figure}
Starting point is an epitaxially grown \CQW\ heterostructure depicted in
Fig.~\ref{fig:indexc}(a). Two GaAs layers with a thickness of 8\,nm each form
the {\QW}s, while the coupling strength is given by a 4\,nm tunnel barrier made
out of Al$_{0.3}$Ga$_{0.7}$As. The center of the \CQW\ structure is located
60\,nm below the surface to assure excellent optical access. At a depth of
370\,nm, an n-doped GaAs layer serves as back gate. In conjunction with
lithographically defined metallic gate structures deposited on the sample, an
electric field parallel to the crystal growth direction can be applied and
spatially varied. As sketched in Fig.~\ref{fig:indexc}(b), the resulting
voltage-tunable tilt of the band structure allows the formation of spatially
indirect excitons (dashed ellipse). Initially photo-generated electrons (open
circle) and holes (black circle) are spatially separated by the tunnel barrier.
The excitonic life-time of typically $\approx$\,1\,ns in bulk material is
extended into the $\mu$s regime, exceeding the excitonic cooling time of
typically $\approx 400$\,ps by far~\cite{DamPRB90}. The energetic red-shift of
the indirect excitons, mediated by the \QCSE, can be tuned via the strength of
the external electric field. Lithographically structured top gate designs allow
to tune the energy of the excitons within the \CQW\ plane in dependence on
position, and potential landscapes for excitons of controlled geometries are
formed. This approach also allows to temporarily control the potential landscape
in-situ via the applied gate voltages.

Time-resolved photoluminescence (\PL) is used to follow the spatial and temporal
decay of excitons. The experiments are carried out in a continuous flow cryostat
at a temperature of $3.8$\,K. The {\CQW}s are selectively populated with
indirect excitons by a pulsed laser with a wavelength of $\lambda=680$\,nm. The
diameter of the laser spot on the sample is $< 20$\,$\mu$m. The delayed \PL\
emission occuring at a wavelength of about $\gtrsim 800$\,nm is detected normal
to the surface via a gated intensified CCD camera. The spatial resolution of
$\approx 1$\,$\mu$m enables to directly reveal the lateral distribution of
excitons. A long-pass filter bocks non-excitonic \PL\ of a wavelength less than
$\approx 780$\,nm. The camera's shutter is set to an exposure time of 50\,ns.
Each experiment is performed at a repetition rate of 100\,kHz and is integrated
for 40\,s in order to yield a comfortable signal-to-noise ratio.
\section*{Shuttling Excitons}
\begin{figure}[t] 
\begin{center}\leavevmode
\includegraphics[width=0.9\linewidth]{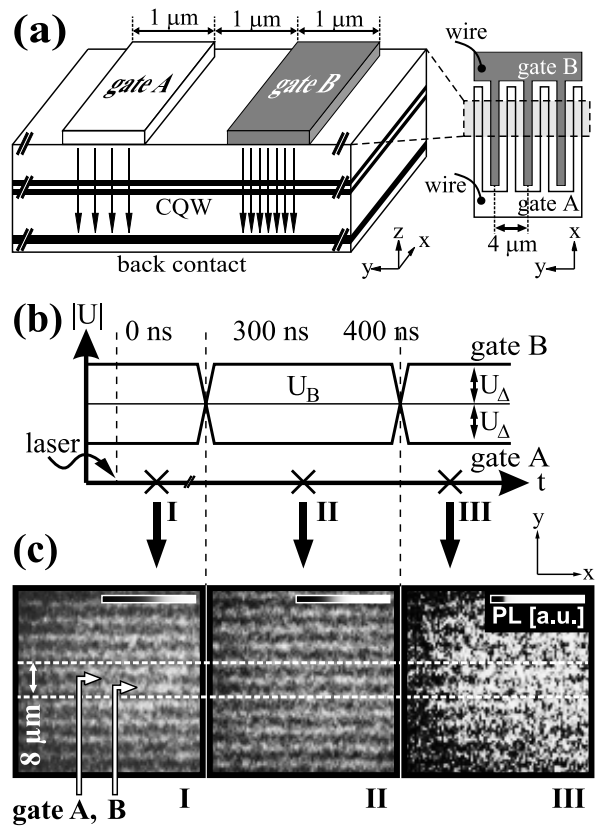}
\end{center}
\caption{
Shuttling of indirect excitons in a time-dependent lateral potential
modulation. 
(a) An interdigitated gate design defines a laterally undulated potential
landscape for excitons. The strength of the electric field is indicated by the
density of vertical arrows. Inset: top view cutout of the gate design.
(b) Chronology of laser excitation ($t=0$\,ns) and the voltage configurations at
times of $t=200$\,ns (I), $t=350$\,ns (II), and $t=440$\,ns (III).
(c) Images of the lateral \PL\ distribution taken at I, II, and III. Long-living
excitons (bright \PL\ lines) are shuttling between the gate fingers along the
y-direction.
}
\label{fig:idgate}
\end{figure}
A semi-transparent interdigitated gate structure with a periodicity of 4\,$\mu$m
is deposited on top of the sample similar to ref.~\cite{ZimAPL98}.
Fig.~\ref{fig:idgate}(a) sketches two adjacent gate fingers labeled \q{gate A}
and \q{gate B}. They are made out of NiCr (10\,nm thickness) and measure a
length of 500\,$\mu$m. Fig.~\ref{fig:idgate}(b) shows the tenor of the
experiment. A bias voltage of $\ub = -450$\,mV and a differential voltage of
$\ud=50$\,mV are applied to the gates and define an undulated lateral potential
landscape for long-living indirect excitons within the plane of the {\CQW}s. The
lateral potential modulation is chosen to be sufficiently small to avoid exciton
ionisation \cite{KraPRL02}. Population of the {\CQW}s with excitons is performed
via subsequent laser illumination for $50$\,ns, with the time $t=0$\,ns marking
the end of the excitation pulse. At a time of $t=200$\,ns after laser
illumination (indicated by \q{I}), the lateral distribution of the emitted \PL\
is imaged by the \ICCD\ camera. The voltages of gate A and gate B are exchanged
at a time of $t=300$\,ns, and a second image (indicated by \q{II}) is taken at a
time of $t=350$\,ns. A second gate voltage reversal follows at $t=400$\,ns, and
a third image (indicated by \q{III}) is taken at $t=440$\,ns. Cutouts of the
image data obtained in this experiment are shown in Fig.~\ref{fig:idgate}(c).
With the position of the gates A und B being indicated, the \PL\ is aligned with
respect to the gate fingers. Being \q{high-field-seekers}, excitons accumulate
underneath the gate of stronger electric field minimizing their potential
energy. 
\begin{figure}[t] 
\begin{center}
\includegraphics[width=0.9\linewidth]{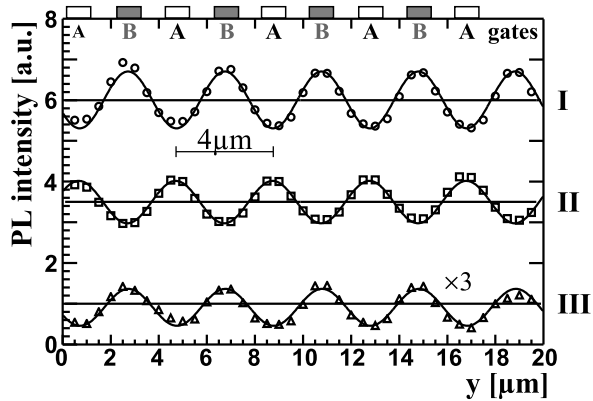}
\end{center}
\caption{
Line-by-line integrated \PL\ yielded from the data shown in
Fig.~\ref{fig:idgate}(c). A constant offset was added to each curve for clarity.
Excitons are collected underneath the gate finger of larger electric field
(maximum in {\PL}-intensity). Excitonic motion is initiated by swapping the gate
polarities (I $\to$ II and II $\to$ III).
}
\label{fig:id-osc}
\end{figure}
A line-by-line integrated analysis of the data is depicted in
Fig.~\ref{fig:id-osc}. The data was corrected for unwanted background light.
Sinusoidal curves (I-III) were fitted to the \PL\ intensity data, corresponding
to the respective images in Fig.~\ref{fig:idgate}(c). In all curves the
4-$\mu$m-periodicity of the interdigitated gate structure is nicely reproduced.
By swapping the gate voltages the repulsive and attractive action of the gate
fingers exchanges. As can be seen in curve II the \PL\ is shifted by 2\,$\mu$m
compared to curve I, indicating that the mobile excitons follow the moving
potential. The second gate voltage reversal (II $\to$ III) completes the
excitonic shuttling process. Regarding the sequence of curve I through curve
III, the \PL\ amplitude is diminishing in agreement with the fact that the
number of excitons decays in time due to recombination.
\section*{Long-range drift}
\begin{figure}[t]
\begin{center}
\includegraphics[width=0.9\linewidth]{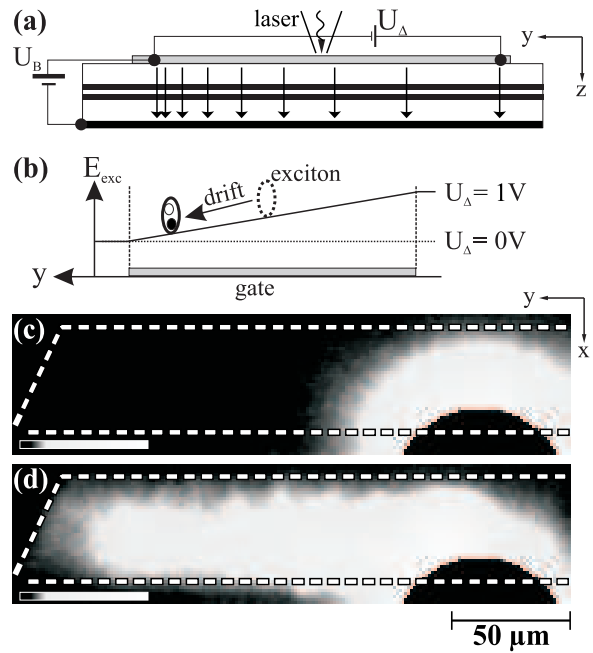}
\end{center}
\caption{
(a) A resisitive gate stripe on top of the sample (grey) is used to define a
linear gradient potential for excitons. The strength of the electric field is
indicated by the density of vertical arrows.
(b) Exciton drifting along the gradient. The slope is tunable via the voltage
difference $\ud$.
(c) Greyscale image of the \PL\ distribution taken with the gradient potential
switched off ($\ud=0$\,V). Excitons are created underneath the black disk
located at the rim of the resistive gate (dashed region). (d) Excitonic drift
over more than $150$\,$\mu$m is observed at a voltage difference of $\ud=+1$\,V.
}
\label{fig:gradpot}
\end{figure}
In order to study long-range excitonic drift a resistive gate stripe was defined
on top of the heterostructure represented by the grey area in
Fig.~\ref{fig:gradpot}(a). The length of the semitransparent titanium gate is
500\,$\mu$m, its width equals 50\,$\mu$m, and its thickness is 10\,nm. A bias
voltage of $\ub=-600$\,mV is applied corresponding to a maximal vertical
electric field of $3.5 \times 10^6$\,V/m at the left side of the gate. This
estimate accounts for an intrinsic bias voltage of $\approx -700$\,mV provided
by the metal/semiconductor interface. An optional voltage difference $\ud$ of
$\pm 1$\,V over the gate stripe can be applied. The resulting strength of the
lateral electric field of $\approx 3 \times 10^3$\,V/m is small compared to the
strength of the vertical electric field. Both are set to temporary constant
values during the experiment. Subsequently, by illuminating the sample by a
laser pulse of a duration of 50\,ns and assisted by the bias voltage $\ub$,
long-living indirect excitons are created. Via the voltage drop $\ud$ over the
resistive gate stripe a gradient potential for excitons can be induced in the
{\CQW}-layer as sketched in Fig.~\ref{fig:gradpot}(b). The slope of the
{\QCSE}-mediated gradient is tunable via the voltage difference $\ud$. The
excitation laser beam was focused to the rim of the gate stripe, located
underneath the black disk shown in Fig.~\ref{fig:gradpot}(c). This configuration
enables to spatially separate mobile excitons in the {\CQW}s from slowly
decaying stationary \PL\ originating from bulk GaAs defects. After a time of
50\,ns following the illumination, a spatially resolved top view image of the
delayed PL is taken by the \ICCD\ camera. Fig.~\ref{fig:gradpot}(c) shows the
experimental result without using a voltage difference ($\ud=0$\,V). No directed
drift is observed as the gradient potential is not swiched on, but a uniform
diffusive excitonic cloud spreads in the vicinity of the excitation spot. In
Fig.~\ref{fig:gradpot}(d), $\ud$ is set to $+1$\,V, exposing the excitons to a
gradient potential as shown in Fig.~\ref{fig:gradpot}(b). Under its influence
the excitons below the gate stripe start to travel along the y-axis towards the
region of stronger vertical electric field. Setting the voltage difference $\ud$
to $-1$\,V reverses the drift direction (not shown). Drift of individual
electrons and holes can be excluded as they would be forced to travel in
opposite directions by the voltage difference $\ud$. Due to the spatial
separation, no recombination PL would occur~\cite{KraPRL02}. It is worth noting
that in contrast to ref.~\cite{HagAPL95} in this experiment the drift covers a
macroscopic distance exceeding 150\,$\mu$m, and is only limited by the length of
the gate stripe. A first estimate on the lower limit of the drift velocity $\vd$
of indirect excitons can be given. The excitons are drifting during a period of
time of $\le 150$\,ns from the beginning of the laser illumination until the end
of the camera exposure. Together with the drift length measured to be $\ge
150$\,$\mu$m, a minimum drift velocity $\vd = 10^3$\,$ \mathrm{m} / \mathrm{s}$
is deduced for this configuration, being comparable to the speed of sound in
GaAs. 
\section*{Summary}
Our experiments demonstrate that voltage-tunable artificial potentials can be
employed to induce excitonic drift over macroscopic distances. This enables us
to design and to test artificial excitonic traps needed to accumulate large
exciton densities, a prerequisite for the observation of \BEC.

We thank J.~Krau\ss\ and A.~W.~Holleitner for valuable discussions as well as
the Deutsche Forschungsgemeinschaft for financial support.

\end{document}